\documentclass[11pt,a4paper]{article}

\usepackage{geometry}
\usepackage{cite}

\usepackage{graphicx}
\usepackage{subfig} 

\usepackage{amsmath,amssymb,amsfonts}

\usepackage{url}
\usepackage{booktabs}
\usepackage{tikz}
\usetikzlibrary{arrows.meta, positioning}
\usepackage{authblk} 
\usepackage{hyperref} 

\usepackage{fancyhdr}
\fancypagestyle{preprint}{
  \fancyhf{}
  
  \fancyfoot[C]{\footnotesize This work has been submitted to the IEEE for possible publication. Copyright may be transferred without notice, after which this version may no longer be accessible.}
}

\usepackage{listings}
\usepackage{xcolor}

\lstdefinestyle{code}{
  basicstyle=\ttfamily\footnotesize,
  breaklines=true,
  frame=single,
  columns=fullflexible,
  aboveskip=8pt,
  belowskip=8pt,
  captionpos=b,       
  framesep=4pt        
}
\begin{document}

% ===================================================================
% TITLE & AUTHORS
% ===================================================================

\title{Enterprise Identity Integration for AI-Assisted Developer Services: Architecture, Implementation, and Case Study}

\author{Manideep Reddy Chinthareddy}
\affil{Senior Software Engineer, Centerville, USA \\ \texttt{chmanideepreddy@gmail.com}}

\date{November, 2025}

\maketitle

% Apply the copyright notice footer to this page only
\thispagestyle{preprint}

% ===================================================================
% ABSTRACT & INDEX TERMS
% ===================================================================

\begin{abstract}
AI-assisted developer services are increasingly embedded in modern IDEs, yet enterprises must ensure these tools operate within existing identity, access control, and governance requirements. The Model Context Protocol (MCP) enables AI assistants to retrieve structured internal context, but its specification provides only a minimal authorization model and lacks guidance on integrating enterprise SSO. This article presents a practical architecture that incorporates OAuth 2.0 and OpenID Connect (OIDC) into MCP-enabled developer environments. It describes how IDE extensions obtain and present tokens, how MCP servers validate them through an identity provider, and how scopes and claims can enforce least-privilege access. A prototype implementation using Visual Studio Code, a Python-based MCP server, and an OIDC-compliant IdP demonstrates feasibility. A case study evaluates authentication latency, token-validation overhead, operational considerations, and AI-specific risks. The approach provides a deployable pattern for organizations adopting AI-assisted developer tools while maintaining identity assurance and auditability.
\end{abstract}

\vspace{0.5cm}

% ===================================================================
% INTRODUCTION
% ===================================================================

\section{Introduction}
\label{sec:introduction}

AI-assisted developer tools such as GitHub Copilot and IDE-native AI assistants have rapidly moved from experimental features to mainstream productivity tools. At the same time, the Model Context Protocol (MCP) has emerged as a standard mechanism for integrating structured, domain-specific context into AI workflows, enabling developer tools to query internal documentation, source code repositories, build systems, and other enterprise resources through MCP servers.

Despite this potential, enterprise adoption remains cautious. Organizations must ensure that AI-assisted interactions respect existing identity, access control, and governance policies. While the 2025 MCP specification introduces a basic authorization model through bearer tokens, it does not prescribe how tokens are issued, how they should be validated, or how they should be mapped to enterprise roles, attributes, and policies. It also does not provide concrete patterns for integrating IDE-based AI assistants—which behave as public clients running on developer machines—with enterprise SSO and least-privilege authorization.

In this article, we address this implementation gap by proposing and demonstrating an architecture that integrates enterprise identity with MCP-enabled AI developer services. The focus is on MCP-specific mechanisms (such as the \texttt{authorization} field and well-known metadata endpoints), on the challenges of managing tokens in IDE environments, and on tailoring OAuth~2.0/OIDC patterns to AI-assisted developer workflows rather than generic web applications.

From a practitioner standpoint, the goal is to provide concrete guidance that teams can apply with any standards-compliant IdP. While the implementation uses Keycloak for illustration, the architecture generalizes to other enterprise IAM platforms such as Azure Active Directory, Okta, or PingFederate.

% -------------------------------------------------------------------
\subsection{Motivation}

Developers increasingly expect AI-assisted experiences in their IDEs, yet enterprises must protect proprietary code, infrastructure details, and internal documentation. Without a well-defined identity integration pattern, MCP servers risk being over-privileged or inconsistently configured, leading to security, compliance, and audit challenges. In addition, AI assistants can act as powerful intermediaries that chain together MCP tools and large language models; if the assistant is bound to an over-privileged identity, it may unintentionally exfiltrate or overexpose sensitive data.

% -------------------------------------------------------------------
\subsection{Contributions}

The main contributions of this article are:

\begin{itemize}
  \item An analysis of the MCP authorization model and its limitations from an enterprise identity perspective, with emphasis on IDE-based AI assistants as public OAuth~2.0 clients.
  \item A reference architecture that integrates MCP servers with OAuth~2.0/OIDC-based enterprise SSO using MCP-specific discovery and authorization mechanisms.
  \item An implementation blueprint using an OIDC-compliant identity provider (illustrated with Keycloak), a Python-based MCP server, and a Visual Studio Code AI extension, together with guidance on how to transfer the pattern to other IdPs.
  \item A case study demonstrating practical integration patterns, latency and token validation characteristics, AI-specific security risks, and operational considerations.
\end{itemize}

Unlike prior work on OAuth/OIDC integration for web and mobile applications, this article is the first to present a comprehensive identity-integration pattern specifically tailored for MCP-enabled AI developer environments. Existing IAM literature does not address how IDE-based AI assistants, which function as public clients orchestrating multiple enterprise resources, should obtain, present, or scope access tokens when interacting with MCP servers. The architecture, implementation blueprint, and case study provided here fill this gap by unifying MCP authorization semantics with enterprise SSO, introducing concrete tooling patterns for token validation, PKCE-based authentication, scope-to-tool mapping, and secure IDE token handling. To our knowledge, no previous work has proposed a deployable, end-to-end identity model that connects MCP servers, OIDC-compliant IdPs, and AI-assisted IDE extensions in a manner suitable for production enterprise environments. 

The remainder of this article is organized as follows. Section~\ref{sec:background} provides background on MCP and enterprise identity. Section~\ref{sec:architecture} describes the proposed architecture. Section~\ref{sec:implementation} presents implementation details. Section~\ref{sec:casestudy} discusses the case study and evaluation. Section~\ref{sec:lessons} summarizes lessons learned, and Section~\ref{sec:conclusion} concludes the article.

% ===================================================================
% BACKGROUND AND CONTEXT
% ===================================================================

\section{Background and Context}
\label{sec:background}

% -------------------------------------------------------------------
\subsection{Model Context Protocol (MCP)}

The Model Context Protocol defines a standardized way for AI clients to interact with tools and data providers known as MCP servers. The recent specification introduces an \texttt{authorization} field that allows clients to present bearer tokens alongside their requests, enabling servers to perform authorization checks according to enterprise policies. MCP also defines discovery mechanisms, including well-known endpoints for OAuth-protected resources, that AI clients can use to learn how to obtain and present access tokens when interacting with MCP servers.

These features make MCP a natural integration point between enterprise IAM and AI-assisted development, but the specification intentionally leaves token issuance, validation strategy, and policy mapping to implementers. As a result, enterprises must define their own patterns for connecting MCP servers to existing OAuth~2.0/OIDC infrastructure.

% -------------------------------------------------------------------
\subsection{Enterprise Identity and Access Management}

Enterprise environments typically rely on identity and access management (IAM) platforms implementing OAuth~2.0 and OpenID Connect (OIDC). These platforms act as identity providers (IdPs), issue tokens, and enforce authentication policies such as multi-factor authentication (MFA) and conditional access. Common IdPs include Keycloak, Azure Active Directory, Okta, PingFederate, and other OIDC-compliant platforms.

Although configuration details differ across products, the underlying building blocks are similar: public and confidential clients, redirect URIs, scopes, claims, JSON Web Tokens (JWTs), and key management via JSON Web Key Sets (JWKS). MCP servers can be treated as OAuth~2.0 resource servers, while IDE extensions and AI assistants behave as public clients that must use Proof Key for Code Exchange (PKCE) and other native-app safeguards.

% -------------------------------------------------------------------
\subsection{AI-Assisted Developer Services in IDEs}

Modern IDEs host AI-assisted extensions that communicate with both model providers and enterprise systems. Integrating identity into this path requires careful handling of tokens in public client environments and ensuring that the AI assistant acts within the bounds of the authenticated developer's permissions.

Unlike browser-based applications, IDE-based clients often run on heterogeneous developer workstations, may not have a traditional backend component, and must manage tokens through local operating-system keychains or encrypted storage. When those clients also integrate MCP, they effectively become orchestrators that can chain multiple MCP tools under a single identity, amplifying the impact of any misconfigured scopes or roles. This makes the design of identity integration patterns particularly important for AI-assisted developer services.

% ===================================================================
% ARCHITECTURE
% ===================================================================

\section{Architecture for Enterprise Identity Integration}
\label{sec:architecture}

In this section, we describe a reference architecture that extends MCP's authorization model with enterprise identity integration in the context of IDE-based AI assistants.

% -------------------------------------------------------------------
\subsection{High-Level Components}

The architecture consists of three main components:
\begin{enumerate}
  \item \textbf{IDE AI Extension}: A plugin in Visual Studio Code (VS Code) or IntelliJ that implements an MCP client and acts as a public OAuth~2.0 client.
  \item \textbf{MCP Server}: A resource server that validates tokens and applies authorization policies to MCP tools and resources.
  \item \textbf{Identity Provider (IdP)}: An OAuth~2.0/OIDC-compliant SSO provider (for example, Keycloak, Azure Active Directory, or Okta) that issues tokens and enforces authentication policies.
\end{enumerate}

% -------------------------------------------------------------------
\subsection{Authentication and Authorization Flow}

The interaction proceeds as follows:
\begin{enumerate}
  \item The IDE extension attempts to invoke an MCP tool without valid credentials.
  \item The MCP server responds with \texttt{401 Unauthorized} and a \texttt{WWW-Authenticate} header indicating the need for a bearer token.
  \item The IDE triggers a browser-based OAuth~2.0/OIDC flow against the enterprise IdP using PKCE.
  \item Upon successful authentication, the IDE obtains an access token and optionally a refresh token.
  \item Subsequent MCP requests include the token in the \texttt{authorization} field defined by the MCP specification.
  \item The MCP server validates the token, evaluates scopes and claims, and authorizes or denies the requested operation.
\end{enumerate}

% Figure 1
\begin{figure}[!t]
\centering
% Adjusted resizebox for standard article textwidth
\resizebox{0.9\textwidth}{!}{%
\begin{tikzpicture}[
    box/.style={rectangle, draw, rounded corners,
                minimum width=2.6cm, minimum height=0.9cm,
                align=center},
    arrow/.style={-Latex, thick},
    node distance=1.6cm and 2.6cm
]

% Top row: IDE, MCP, IdP
\node[box] (ide) {IDE AI Extension\\(MCP Client)};
\node[box, right=of ide] (mcp) {MCP Server};
\node[box, right=of mcp] (idp) {Identity Provider\\(OAuth 2.0 / OIDC)};

% Bottom: Enterprise resources
\node[box, below=1.8cm of mcp] (resources) {Enterprise Resources\\(Code Search, Docs, Build Systems)};

% IDE <-> MCP
\draw[arrow] (ide) -- node[above, yshift=9pt]{MCP requests / responses} (mcp);

% MCP -> IdP (token validation)
\draw[arrow] (mcp) -- node[above, xshift=40pt,yshift=-30pt]{Token validation (JWKS/introspection)} (idp);

% MCP -> resources
\draw[arrow] (mcp) -- node[right, xshift=2pt,yshift=-15pt]{Tool invocations} (resources);

% IDE -> IdP (OAuth authorization, curved)
\path (ide.north) ++(0,0.25) coordinate (startA);
\path (idp.north) ++(0,0.25) coordinate (endA);
\draw[arrow]
  (startA) to[bend left=20]
  node[above, yshift=2pt]{OAuth authorization (browser-based)}
  (endA);

\end{tikzpicture}}
\caption{High-level architecture for enterprise identity integration with MCP servers, IDE AI extensions, and OAuth/OIDC identity providers.}
\label{fig:architecture}
\end{figure}
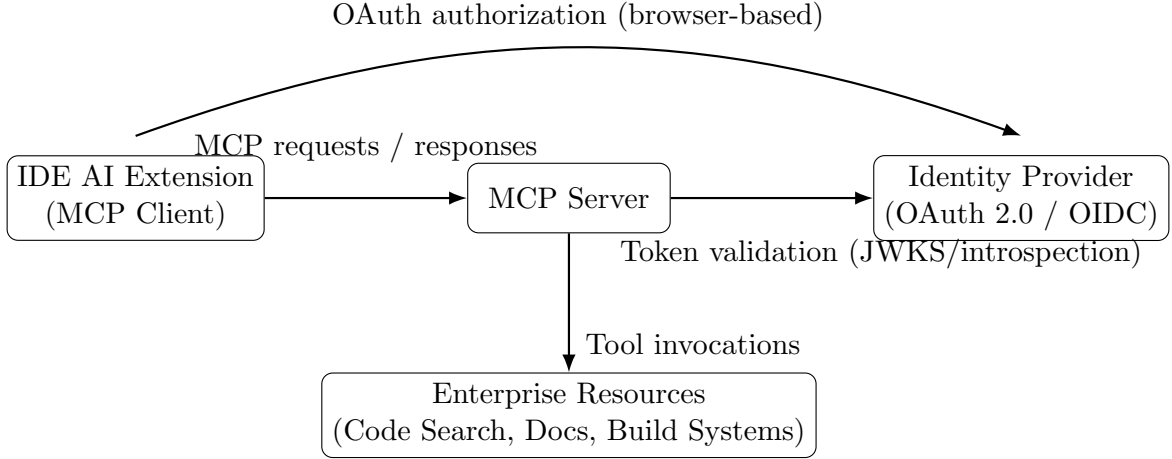

% Figure 2
\begin{figure}[!t]
\centering
% Adjusted resizebox for standard article textwidth
\resizebox{0.9\textwidth}{!}{%
\begin{tikzpicture}[
    actor/.style={rectangle, draw, rounded corners,
                  minimum width=3.2cm, minimum height=1.9cm,
                  align=center},
    lifeline/.style={dashed},
    arrow/.style={-Latex},
    node distance=3.2cm
]

% Actors
\node[actor] (client) {Client\\(MCP IDE Extension)};
\node[actor, right=of client] (mcp) {MCP Server\\(Resource Server)};
\node[actor, right=of mcp] (authz) {Authorization Server\\(OAuth 2.0 / OIDC)};

% Lifelines
\draw[lifeline] (client.south) -- ++(0,-12);
\draw[lifeline] (mcp.south)    -- ++(0,-12);
\draw[lifeline] (authz.south)  -- ++(0,-12);

% compact vertical positions
\def\A{-0.8}
\def\B{-1.8}
\def\C{-2.8}
\def\D{-3.8}
\def\E{-4.8}
\def\F{-5.8}
\def\G{-6.8}
\def\H{-7.8}
\def\I{-8.8}
\def\J{-9.8}
\def\K{-10.8}
\def\L{-11.8}
\def\M{-12.8}

% 1. Initial unauthenticated request
\draw[arrow] ([yshift=\A cm]client.south) --
             ([yshift=\A cm]mcp.south)
    node[midway, above] {1. MCP request without token};

\draw[arrow] ([yshift=\B cm]mcp.south) --
             ([yshift=\B cm]client.south)
    node[midway, above] {2. 401 Unauthorized (may include WWW-Authenticate)};

% 2. Metadata via header
\draw[arrow] ([yshift=\C cm]client.south) --
             ([yshift=\C cm]mcp.south)
    node[midway, above] {3. GET \texttt{resource\_metadata} URI};

\draw[arrow] ([yshift=\D cm]mcp.south) --
             ([yshift=\D cm]client.south)
    node[midway, above] {4. Resource metadata with authorization server URL};

% 3. Fallback to well-known URI
\draw[arrow] ([yshift=\E cm]client.south) --
             ([yshift=\E cm]mcp.south)
    node[midway, above] {5. GET \texttt{/.well-known/oauth-protected-resource/mcp}};

\draw[arrow] ([yshift=\F cm]mcp.south) --
             ([yshift=\F cm]client.south)
    node[midway, above] {6. Resource metadata with authorization server URL};

% 4. OAuth Authorization Request (PKCE)
\draw[arrow] ([yshift=\G cm]client.south) --
             ([yshift=\G cm]authz.south)
    node[midway, above] {7. Authorization request (PKCE)};

% Token request
\draw[arrow] ([yshift=\H cm]client.south) --
             ([yshift=\H cm]authz.south)
    node[midway, above] {8. Token request (PKCE)};

% Access token
\draw[arrow] ([yshift=\I cm]authz.south) --
             ([yshift=\I cm]client.south)
    node[midway, above] {9. Access token (and claims)};

% 5. Protected resource access
\draw[arrow] ([yshift=\J cm]client.south) --
             ([yshift=\J cm]mcp.south)
    node[midway, above] {10. MCP request + Bearer token};

% Token validation
\draw[arrow] ([yshift=\K cm]mcp.south) --
             ([yshift=\K cm]authz.south)
    node[midway, above] {11. Token validation (JWKS / introspection)};

% Validation result
\draw[arrow] ([yshift=\L cm]authz.south) --
             ([yshift=\L cm]mcp.south)
    node[midway, above] {12. Validation result};

% Authorized output
\draw[arrow] ([yshift=\M cm]mcp.south) --
             ([yshift=\M cm]client.south)
    node[midway, above] {13. Authorized MCP response};

\end{tikzpicture}}
\caption{End-to-end MCP authorization sequence including metadata discovery, OAuth token issuance, token validation, and protected resource access~\cite{mcp-authorization}.}
\label{fig:mcp-sequence}
\end{figure}
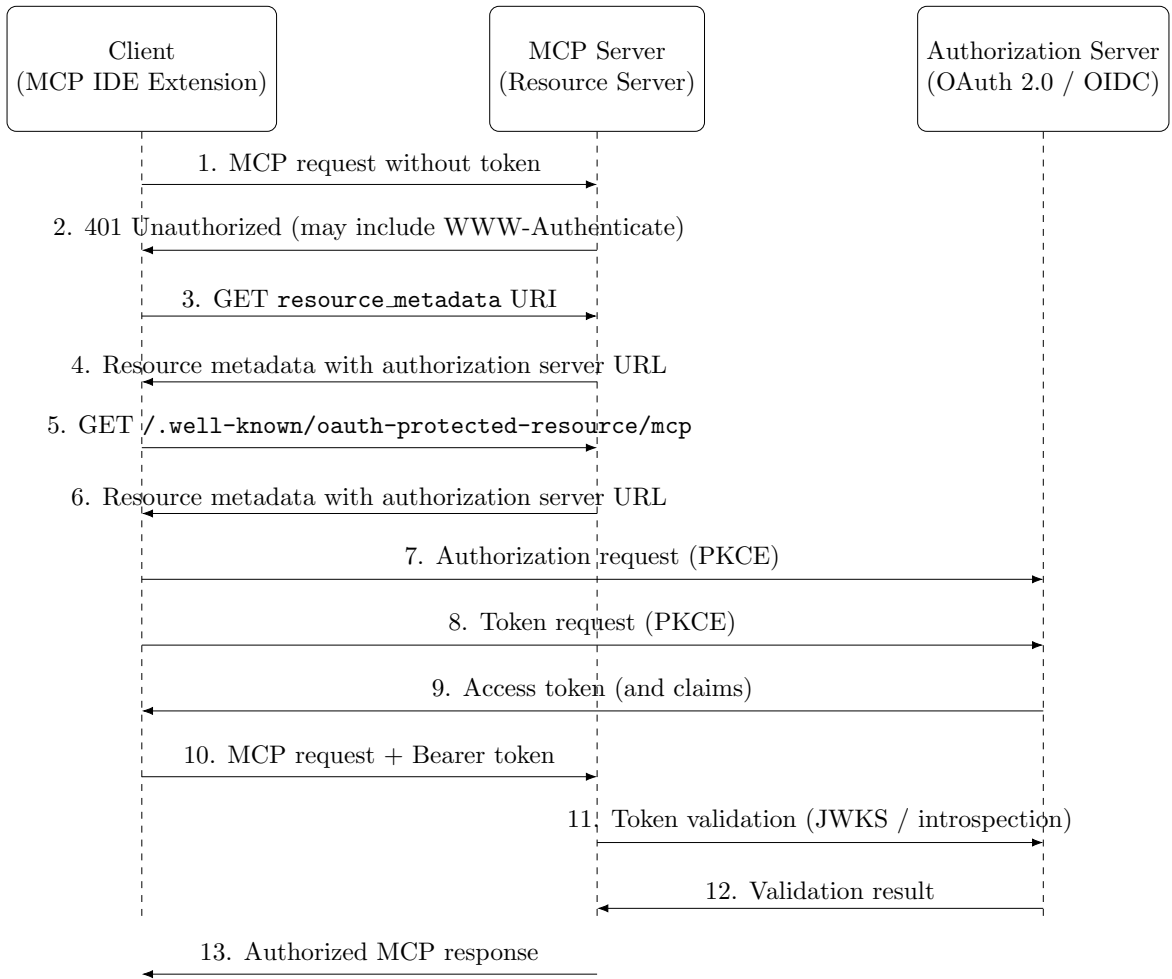

\begin{figure}[!t]
\centering
\subfloat[IdP login page]{%
  \includegraphics[width=0.48\linewidth]{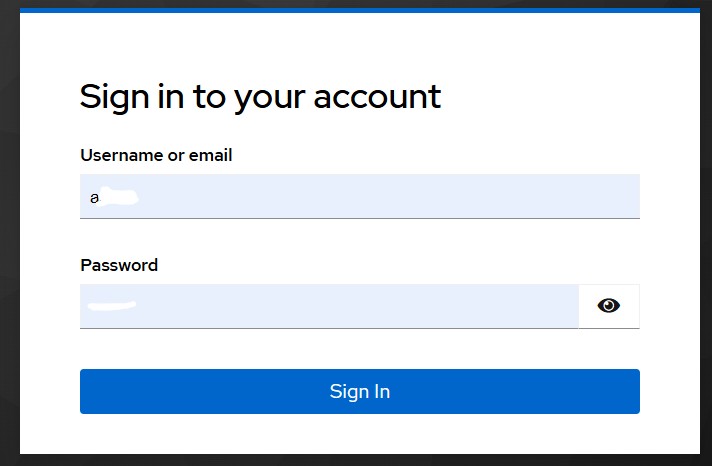}%
}\hfil
\subfloat[Successful login confirmation]{%
  \includegraphics[width=0.48\linewidth]{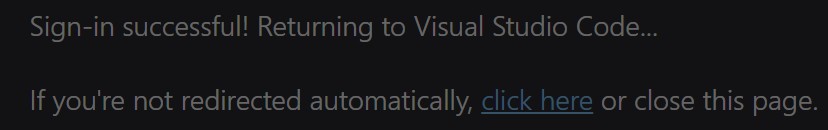}%
}\\[2pt]
\subfloat[VS Code consent prompt]{%
  \includegraphics[width=0.48\linewidth]{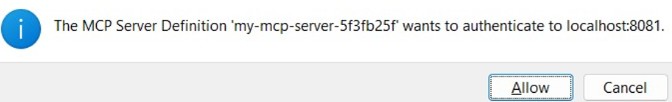}%
}\hfil
\subfloat[Manual client registration prompt]{%
  \includegraphics[width=0.48\linewidth]{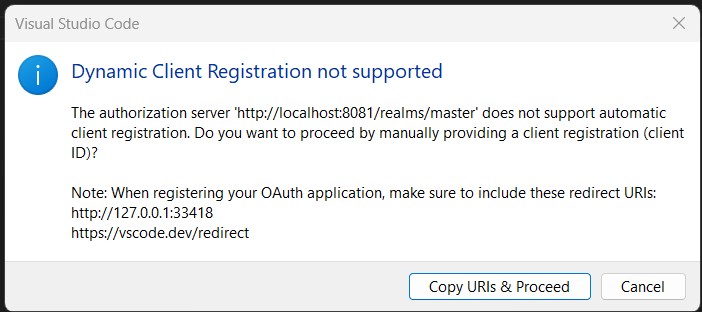}%
}\\[2pt]
\subfloat[Redirect URI configuration in IdP]{%
  \includegraphics[width=0.9\linewidth]{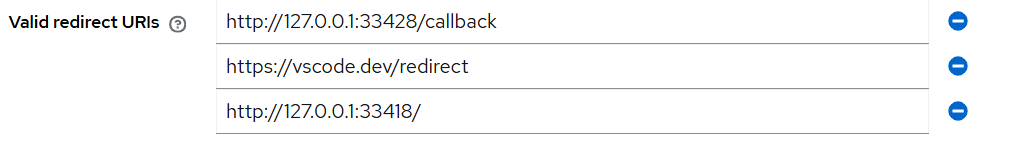}%
}
\caption{Representative OAuth/OIDC experience for developers and administrators when VS Code authenticates an MCP server against an enterprise identity provider.}
\label{fig:oauth-ux}
\end{figure}

Figure~\ref{fig:oauth-ux} illustrates a representative user experience during OAuth/OIDC authentication and client registration when VS Code interacts with an enterprise IdP. After the user successfully authenticates in the browser, VS Code requests consent to authenticate the MCP server against the IdP and, if necessary, guides the user through manual client registration.

% -------------------------------------------------------------------
\subsection{Authorization Models}

Enterprises can implement role-based access control (RBAC), attribute-based access control (ABAC), or scope-based policies that map token claims to permitted MCP tools and resources. For example, a \texttt{developer} role may be allowed to invoke documentation and code-search tools, while a \texttt{contractor} role is restricted to documentation only.

Table~\ref{tab:roles} illustrates a simple mapping from roles and scopes to allowed MCP tools.

\begin{table}[!t]
\centering
\caption{Example role and scope mapping to MCP tools.}
\label{tab:roles}
\begin{tabular}{@{}lll@{}}
\toprule
Role / claim       & Scopes                       & Allowed tools                  \\
\midrule
\texttt{developer} & \texttt{mcp.docs.read},      & Docs search, code search,      \\
                   & \texttt{mcp.code.search}     & build system queries           \\
\texttt{contractor}& \texttt{mcp.docs.read}       & Docs search only               \\
\texttt{operator}  & \texttt{mcp.ops.read}        & Deployment and status queries  \\
\bottomrule
\end{tabular}
\end{table}

\subsection{Threat Model and Limitations}

The proposed architecture assumes that:

\begin{itemize}
  \item IDE extensions behave as honest-but-curious public OAuth~2.0 clients without client secrets.
  \item Access tokens are stored securely using operating system keychain facilities where available.
  \item Network channels between IDEs, MCP servers, and IdPs are protected using TLS.
\end{itemize}

If an IDE environment is compromised, an attacker may obtain short-lived access tokens and invoke MCP tools within the limits of the associated scopes and roles. To limit impact, enterprises should prefer short token lifetimes, restrict refresh token issuance, and enforce fine-grained scopes. If an MCP server is compromised, access to downstream enterprise resources is constrained by its own service credentials and by token-verification policies, but runtime audit logs remain critical for incident response.

Beyond traditional OAuth threat scenarios, AI-assisted developer tools introduce additional considerations. Because the AI assistant can autonomously chain MCP tool invocations based on natural-language prompts, any over-privileged scope configuration can translate into broad, unanticipated access to code, documentation, or operational data. To mitigate this risk, MCP tools should be grouped and scoped along clear data-boundary and role lines, and sensitive tools (for example, those that access production configuration or customer data) should require distinct scopes that are rarely granted to day-to-day developer roles.

These limitations and risks are typical of OAuth-based architectures extended to AI assistants and must be accounted for in enterprise risk assessments.

Recent work on AI-assisted developer environments highlights the operational and governance risks that arise when tools autonomously chain privileged actions or handle sensitive internal context~\cite{Human-AI-Interaction}, ~\cite{mcp-security-framework}, ~\cite{mcp-governance}. Fang et al. demonstrated that LLM agents can autonomously exploit real-world vulnerabilities when provided with tool access~\cite{fang2024}, underscoring the need for strict access boundaries. Additionally, while runtime approaches like AgentBound~\cite{agentbound} focus on execution policies, our architecture emphasizes upstream identity assurance to prevent unauthorized tool access initially.

% ===================================================================
% IMPLEMENTATION DETAILS
% ===================================================================

\section{Implementation}
\label{sec:implementation}

This section outlines a practical implementation using an OIDC-compliant IdP (illustrated with Keycloak as an example), a FastAPI-based MCP server, and a VS Code AI extension. While specific screenshots and configuration names use Keycloak, the same pattern applies to other IdPs with minor adjustments to claims and client-registration workflows.

% -------------------------------------------------------------------
\subsection{Identity Provider Configuration}

The identity provider is configured with a dedicated realm or application for developer tools, a public client representing the IDE extension, and roles or groups corresponding to developer, operator, and contractor personas. Redirect URLs include the local HTTP endpoint used by VS Code as well as a fallback HTTPS redirect. In IdPs other than Keycloak, the equivalent configuration typically involves registering a public client, enabling PKCE, and adding the IDE's redirect URIs.

% -------------------------------------------------------------------
\subsection{MCP Server Implementation}

The MCP server uses middleware to extract the token from the \texttt{authorization} field, validate it using the IdP's JWKS endpoint, and map claims to internal authorization rules for each MCP tool. Listing~\ref{lst:mcp-app-init} shows a simple setup of an MCP server behind an authentication wall. This setup ensures that the MCP server responds to the MCP client with specific headers, informing the client that it is not authenticated and indicating how to authenticate.

\begin{lstlisting}[style=code,float=!t,caption={FastMCP app initialization with OIDC-based authorization.},label={lst:mcp-app-init}]
app = FastMCP(
    "Sample MCP server",
    port=8000,
    token_verifier=RoleAwareTokenVerifier(),
    auth=AuthSettings(
        issuer_url="http://localhost:8081/realms/master",
        resource_server_url="http://localhost:8000/mcp",
        required_scopes=["openid", "profile"]
    )
)
\end{lstlisting}

The \texttt{RoleAwareTokenVerifier} in Listing~\ref{lst:mcp-app-init} is a straightforward implementation of a token verifier that validates JWTs using the IdP's keys and maps claims to internal roles. Its logic follows standard patterns for OAuth~2.0 resource servers and is omitted for brevity.

The code in Listing~\ref{lst:mcp-oauth-resource} defines an OAuth-protected resource descriptor. The MCP client (in this case, VS Code) retrieves this document to discover the authorization servers that can issue tokens for the MCP resource. The endpoint URL must match the MCP specification exactly, including case sensitivity.

\begin{lstlisting}[style=code,float=!t,caption={OAuth-protected resource descriptor advertised by the MCP server.},label={lst:mcp-oauth-resource}]
fastapi = app.streamable_http_app()
fastapi_lifespan = FastAPI(lifespan=lifespan)

@fastapi_lifespan.get("/.well-known/oauth-protected-resource")
async def oauth_protected_resource():
    return JSONResponse({
        "resource": "http://localhost:8000/mcp",
        "scopes_supported": ["openid", "profile"],
        "authorization_servers": ["http://localhost:8081/realms/master"],
        "bearer_methods_supported": ["header","body"],
    })
\end{lstlisting}

If tool-level scope configuration is required, the configuration in Listing~\ref{lst:mcp_tool_scope} illustrates how to enforce scope requirements for a specific MCP tool.

\begin{lstlisting}[style=code,float=!t,caption={Configuring tool-level scope-based access.},label={lst:mcp_tool_scope}]
from fastapi import Depends, Security
from fastapi.security import HTTPBearer, HTTPAuthorizationCredentials

security = HTTPBearer()

@app.tool()
def mcp_get(access_token: AccessToken = Depends(
        require_scopes("mcp.docs.read"))):
    ...
\end{lstlisting}

Listing~\ref{lst:mcp-auth} shows an excerpt from the server logs during a successful authentication and tool invocation.

\begin{lstlisting}[style=code,float=!t,caption={Example MCP authentication logs.},label={lst:mcp-auth}]
INFO  "POST /mcp HTTP/1.1" 401 Unauthorized
INFO  "GET /.well-known/oauth-protected-resource HTTP/1.1" 200 OK
Verifying token...
Authenticated user: a*************
INFO  "POST /mcp HTTP/1.1" 202 Accepted
INFO  "POST /mcp HTTP/1.1" 200 OK
\end{lstlisting}

% -------------------------------------------------------------------
\subsection{VS Code / IDE Plugin Behavior}

The IDE extension initiates the OAuth flow when receiving an unauthorized response from the MCP server, securely stores tokens, and attaches them to subsequent MCP requests. After authentication completes, the extension can successfully enumerate and invoke MCP tools. In practice, this behavior is largely transparent to the developer after the initial login and consent.

% ===================================================================
% CASE STUDY AND EVALUATION
% ===================================================================

\section{Case Study: Securing MCP-Based AI Developer Services}
\label{sec:casestudy}

This case study evaluates the proposed identity integration architecture in a representative enterprise environment. The objective was to validate (1) security posture, (2) practical performance, and (3) the overall impact on developer workflows when enabling MCP-based AI-assisted development tools.

\subsection{Scenario and Environment}

The evaluation environment consisted of the following components:

\begin{itemize}
    \item \textbf{Identity Provider (IdP):} An OAuth~2.0/OIDC server initially deployed on a local laptop for demonstration, and later on an AWS ECS cluster, enforcing MFA and conditional access policies. Keycloak was used in this implementation, but the same pattern can be reproduced with other OIDC-compliant IdPs.
    \item \textbf{MCP Server:} A FastAPI-based resource server running locally behind a reverse proxy and then containerized and deployed to AWS ECS.
    \item \textbf{IDE Client:} A VS Code extension, in this example the GitHub Copilot extension, implementing the MCP client and acting as a public OAuth~2.0 client using PKCE.
    \item \textbf{Enterprise Resources:} Internal documentation indices, architectural standards metadata, code catalog search tools, and build system queries exposed through MCP tools.
\end{itemize}

Latency measurements were collected using server-side timestamps around token validation and MCP tool invocation calls, averaged across multiple runs under normal network conditions.

\subsection{Authentication Behavior and Developer Impact}

A key focus of the case study was determining whether OAuth/OIDC authentication introduced any material friction or latency to developers.

In practice, authentication latency was not a significant concern. Because MCP-enabled tools are accessed almost exclusively by internal developers, the overall authentication load is modest and easily handled by the enterprise IAM infrastructure. Developers typically authenticate once at the beginning of a session, after which cached credentials are reused for all subsequent MCP interactions.

Even in cases where the initial OAuth flow incurred a minor delay due to browser redirects or MFA prompts, this overhead occurred only once per session and had no sustained impact on usability. The improved identity assurance, centralized governance, and traceability provided by OAuth/OIDC outweighed the small time cost.

\subsection{Token Validation and Runtime Performance}

Once authenticated, the MCP server validates tokens using the IdP's JWKS endpoint. Performance measurements collected during the case study showed:

\begin{itemize}
    \item \textbf{JWKS cache hit latency:} about 5\,ms per validation (almost instantaneous).
    \item \textbf{JWKS cache miss latency:} 25--35\,ms (for the initial key fetch only).
    \item \textbf{MCP tool invocation latency:} 90--120\,ms end-to-end, dominated by resource processing and AI-assisted analysis rather than authentication.
\end{itemize}

These results confirm that token validation introduces negligible overhead and does not negatively affect developer workflow responsiveness.

\subsection{Security and Operational Observations}

During evaluation, several important security and operational insights emerged:

\begin{itemize}
    \item \textbf{PKCE is essential} for securing public clients in IDE environments.
    \item \textbf{Scope-based authorization} (for example, \texttt{mcp.docs.read}, \texttt{mcp.code.search}) provides granular control aligned with enterprise least-privilege policies.
    \item \textbf{Refresh token usage} was deliberately restricted; short-lived access tokens reduce the risk of credential theft.
    \item \textbf{Audit logging} of tool invocations, along with authenticated user identity, significantly improves incident investigation and compliance reporting.
    \item \textbf{Client registration friction} was the most common onboarding issue, especially when dynamic registration was unsupported.
    \item \textbf{Redirect URL whitelisting} was a frequent configuration gap, as VS Code uses local callback URLs that must be explicitly added to the IdP.
    \item \textbf{AI-specific risk management} requires mapping particularly sensitive MCP tools (for example, those that expose production operations or confidential datasets) to dedicated scopes and roles that are constrained to a small set of users.
\end{itemize}

Collectively, these findings demonstrate the importance of coordination among platform engineering, IAM, and developer experience teams during rollout.

\subsection{Summary}

The case study confirms that enterprise identity integration for MCP-based AI developer tools is both feasible and operationally lightweight. OAuth/OIDC authentication introduces negligible overhead and provides meaningful improvements in identity assurance, access governance, and auditability. These results support the viability of securely deploying MCP-enabled developer workflows in production enterprise environments.

% ===================================================================
% LESSONS LEARNED
% ===================================================================

\section{Lessons Learned}
\label{sec:lessons}

Key lessons include the importance of clear scope design, robust token validation, secure storage of credentials in IDE environments, and collaboration between platform, security, and developer experience teams. From a practitioner standpoint, organizations planning a similar deployment should:

\begin{itemize}
  \item Enable PKCE for all IDE-based OAuth~2.0 clients.
  \item Design fine-grained scopes (for example, \texttt{mcp.docs.read}, \texttt{mcp.code.search}) mapped to MCP tools. If only one design decision can be prioritized, getting scope design right up front is the most effective way to prevent over-privileged AI assistants.
  \item Prefer short-lived access tokens and restrict the use of refresh tokens.
  \item Whitelist local IDE redirect URLs in the IdP during client registration and document them clearly for developers. Although Keycloak is used for illustration, the integration pattern is fully portable to any OIDC-compliant enterprise provider. Platforms such as Azure Active Directory, Okta, and PingFederate follow the same underlying OAuth 2.0 and OIDC semantics—public-client registration, PKCE enforcement, JWKS-based token validation, and scope-to-resource mapping—making the architecture directly transferable without modification to protocol logic. 
  \item Integrate MCP tool invocation logs with existing security monitoring and audit pipelines, including attribution to specific users and scopes.
\end{itemize}

Providing visible feedback in both the IDE and server logs helps operators diagnose configuration issues quickly and reduces the time to resolve onboarding problems.

% ===================================================================
% CONCLUSION
% ===================================================================

\section{Conclusion}
\label{sec:conclusion}

MCP-enabled AI developer tools offer powerful new capabilities for accessing enterprise context directly from IDEs. The MCP authorization model provides a necessary foundation, but enterprises must integrate it with their identity infrastructure to meet real-world security and governance requirements.

This article presented an architecture that combines OAuth~2.0/OIDC-based SSO, MCP servers, and IDE extensions into a coherent, secure solution suitable for production environments. By treating IDE-based AI assistants as public OAuth~2.0 clients, leveraging MCP's authorization and discovery mechanisms, and applying scope- and role-based policies to MCP tools, organizations can safely expose internal developer resources to AI workflows.

The implementation patterns and case study results show that the required components can be deployed with modest operational effort and negligible runtime overhead. Practitioners can use the patterns described here---independent of any specific IdP vendor---to accelerate adoption of MCP-enabled AI developer services while maintaining identity assurance, least privilege, and auditability across their software delivery environments.

% ===================================================================
% REFERENCES
% ===================================================================

\end{document}